\documentclass[12pt]{article}

\usepackage{latexsym}
\usepackage{amssymb}
\usepackage{amsmath}

\newtheorem{theorem}{Theorem}

\def\cl{{\mathcal C}\!\ell}
\def\diag{{\rm diag}}
\def\Even{{\rm Even}}
\def\Odd{{\rm Odd}}
\def\Tr{{\rm Tr}}

\def\Spin{{\rm Spin}}
\def\extexp{\widehat{\rm exp}}
\def\R{{\mathbb R}}

\def\proof{\medbreak\noindent{\bf Proof}}
\newcommand{\fin}{\hbox{$\blacksquare$}\medskip\par}

\def\SO{{\rm SO}}

\begin{document}

\title{Parametrisations of elements of spinor and orthogonal groups using
 exterior exponents}

\author{Nikolay Marchuk}

\maketitle

\thanks{This work was partially supported by the grant of the President
of the Russian Federation (project NSh-3224.2008.1) and by
Division of mathematics of RAS (project ``Modern problems in
theoretical mathematics'').}

\begin{abstract}
We present new parametrizations of elements of spinor and
orthogonal groups of dimension 4 using Grassmann exterior algebra.
Theory of spinor groups is an important tool in theoretical and
mathematical physics namely in the Dirac equation for an electron.
\end{abstract}

 \noindent Steklov Mathematical Institute,\\ Gubkina st.8, Moscow 119991, Russia
\medskip

\noindent Email: nmarchuk2005@yandex.ru, nmarchuk@mi.ras.ru

\medskip


\bigskip

An exterior algebra, invented by G.~Grassmann in the year 1844
\cite{Grassmann}, has many applications in different fields of
mathematics and physics. Here we present a new application of
Grassmann algebra to the theory of spinor and orthogonal groups.
\medskip

\noindent {\bf Clifford algebras.} Let $p,q,n$ be nonnegative
integer numbers and $n=p+q$. And let $\cl(p,q)$ be a real Clifford
algebra \cite{mybook} of the signature $(p,q)$ with generators
$e^1,\ldots,e^n$ such that
$$
e^a e^b + e^b e^a = 2\eta^{ab}e,\quad a,b=1,\ldots,n,
$$
where $e$ is the identity element of Clifford algebra and
$\eta^{ab}$ are elements of the diagonal matrix of dimension $n$
$$
\eta=\diag(1,\ldots,1,-1,\ldots,-1)
$$
with $p$ pieces of $1$ and $q$ pieces of $-1$ on the diagonal. The
Clifford algebra $\cl(p,q)$ can be considered as $2^n$-dimensional
vector space with basis elements
\begin{equation}
e,e^a,e^{a_1a_2},\ldots,e^{a_1\ldots a_{n-1}},e^{12\ldots n},\quad
0\leq a_1<\ldots<a_k\leq n \label{cl:basis}
\end{equation}
numbered by ordered multi-indices of lengths from $0$ to $n$. Any
element of Clifford algebra $\cl(p,q)$ can be written in the form
of decomposition w.r.t. the basis (\ref{cl:basis})
\begin{equation}
U=u e +u_a
e^a+\sum_{a_1<a_2}u_{a_1a_2}e^{a_1a_2}+\ldots+u_{1\ldots
n}e^{1\ldots n}\label{decomp}
\end{equation}
with real coefficients $u,u_a,u_{a_1a_2},\ldots,u_{1\ldots n}$.
Elements of the form
$$
U=\sum_{a_1<\ldots<a_k}u_{a_1\ldots a_k}e^{a_1\ldots a_k}
$$
are called {\em elements of rank $k$}. Denote by $\cl_k(p,q)$ the
subspace of rank $k$ elements. We have
$$
\cl(p,q)=\cl_0(p,q)\oplus\ldots\oplus\cl_n(p,q).
$$
An element $U\in\cl(p,q)$ is called {\em even (odd)} if this
element is a sum of elements of even (odd) ranks. Hence
\begin{eqnarray*}
\cl_\Even(p,q)&=&\cl_0(p,q)\oplus\cl_2(p,q)\oplus\ldots,\\
\cl_\Odd(p,q)&=&\cl_1(p,q)\oplus\cl_3(p,q)\oplus\ldots,\\
\cl(p,q)&=&\cl_\Even(p,q)\oplus\cl_\Odd(p,q).
\end{eqnarray*}

\medskip

\noindent{\bf Reverse operation.} Let us define a linear {\em
reverse} operation $\sim : \cl(p,q)\to\cl(p,q)$ with the aid of
the following rules:
$$
e^\sim= e,\quad (e^a)^\sim=e^a,\quad (U V)^\sim=V^\sim
U^\sim,\quad\forall U,V\in\cl(p,q).
$$
In particular,
$$
(e^{a_1}\ldots e^{a_k})^\sim=e^{a_k}\ldots e^{a_1}.
$$
\medskip

\noindent{\bf Spinor groups.} Let $n=p+q\leq 5$. Consider the
following set of even elements of the Clifford algebra $\cl(p,q)$:
$$
\Spin_+(p,q)=\{S\in\cl_\Even(p,q) : S^\sim S=e\}.
$$
This set is closed w.r.t. Clifford product and contains the
identity element $e$. Elements of this set are invertible.
Therefore $\Spin_+(p,q)$ can be considered as a group (Lie group)
w.r.t. the Clifford product. This group is called {\em spinor
group}.

The set of second rank elements $\cl_2(p,q)$ with the commutator
$[A,B]=AB-BA$ is the Lie algebra of the Lie group $\Spin_+(p,q)$
\cite{Lounesto}.

Let us define {\em an exponent} of Clifford algebra elements $\exp
: \cl(p,q)\to\cl(p,q)$ by the formula
$$
\exp\,A=e+A+\frac{1}{2!}A^2+\frac{1}{3!}A^3+\ldots.
$$
\medskip

\noindent{\bf Spinor groups in cases $p+q=4$.} The sign $\simeq$
denotes group isomorphisms. It is known \cite{Lounesto} that
\begin{itemize}
\item $\Spin_+(4,0)\simeq\Spin_+(0,4)\simeq\Spin(4)$. \item
$\Spin_+(1,3)\simeq\Spin_+(3,1)$. \item $\forall S\in\Spin(4)$
there exists $B\in\cl_2(4,0)$ such that $S=\exp\,B$. \item
$\forall S\in\Spin_+(1,3)$ there exists $B\in\cl_2(1,3)$ such that
$S=\exp\,B$, or $S=-\exp\,B$. Note that the set of exponents of
rank 2 elements do not form a group. \item There exist elements
$S\in\Spin_+(2,2)$ such that these elements can't be represented
in the form $\pm\exp\,B$, where $B\in\cl_2(2,2)$.

\end{itemize}

\medskip

\noindent{\bf Exterior (Grassmann) multiplication of Clifford
algebra elements.} Let us define the associative and distributive
operation of exterior multiplication of Clifford algebra elements
(denoted by $\wedge$)
$$
e^{a_1}\wedge e^{a_2}\wedge\ldots\wedge
e^{a_k}=e^{[a_1}e^{a_2}\ldots e^{a_k]},
$$
where square brackets denote the operation of alternation of
indices. The Clifford algebra $\cl(p,q)$, considered with the
exterior product, can be identified with the Grassmann algebra of
dimension $n$.
\medskip

\noindent{\bf Exterior exponent.} Consider the exterior exponent
$\extexp : \cl(p,q) \to \cl(p,q)$
\begin{equation}
\extexp(B)=e+B+\frac{1}{2!} B\wedge B+\frac{1}{3!} B\wedge B\wedge
B +\ldots, \label{extexp}
\end{equation}
where $B\in\cl(p,q)$. We interested in exterior exponent of second
rank elements. In this case in the right hand part of
(\ref{extexp}) there are finite number of nonzero summands. In
particular, for the case  $n=4$ there are only three summands
$$
\extexp(B)=e+B+\frac{1}{2} B\wedge B,
$$
where $B\in\cl_2(p,q)$, $p+q=4$. We see that
$$
(\extexp(B))^\sim=\extexp(-B)=e-B+\frac{1}{2} B\wedge B.
$$
It is not hard to prove that
$$
(\extexp(B))^\sim \extexp(B)=\lambda e,
$$
where $\lambda=\lambda(B)$ is a scalar that depends on
coefficients of the element $B$.

\medskip

\noindent{\bf Main theorem.} If $U\in\cl(p,q)$ is written in the
form (\ref{decomp}), then we denote $\Tr\,U=u$. Also denote
$\ell=e^1e^2e^3e^4$ and $\epsilon=\Tr(\ell^2)$.

\begin{theorem}
Let $n=p+q=4$. Any element $S$ of the group $\Spin_+(p,q)$ can be
represented in one of two following forms:
\begin{itemize}
\item If $\Tr\,S\neq 0$, then there exists $B\in\cl_2(p,q)$ such
that $\lambda=\lambda(B)>0$ and
\begin{equation}
S=\pm\frac{1}{\sqrt{\lambda}}\extexp\,B.\label{exex}
\end{equation}
The sign (plus or minus) at the right hand part is equal to the
sign of the number $\Tr\,S$. \item If $\Tr\,S=0$, then there
exists $B\in\cl_2(p,q)$ such that $B\wedge B=0$,
$\epsilon(1+\beta)\geq 0$, where $\beta=\Tr(B^2)$ and
\begin{equation}
S=B\pm\ell\sqrt{\epsilon(1+\beta)}.\label{S:Badj}
\end{equation}
The sign (plus or minus) at the right hand part is equal to the
sign of the number $\Tr(\ell^{-1}S)$.
\end{itemize}

\end{theorem}

\proof. Let $S\in\Spin_+(p,q)$ be such that $\Tr\,S=\alpha\neq 0$.
We white $S$ and $S^\sim$ in the form
$$
S=\alpha e+U+F,\quad S^\sim=\alpha e-U+F,
$$
where $U\in\cl_2(p,q)$, $F\in\cl_4(p,q)$. For $n=4$ it follows
that
$$
U^2=U\wedge U+\gamma e,\quad \alpha^2 e-F^2-\gamma^2
e\in\cl_0(p,q),\quad 2\alpha F-U\wedge U\in\cl_4(p,q).
$$
The identity
$$
S^\sim S=(\alpha^2 e-F^2-\gamma^2 e)+(2\alpha F-U\wedge U)=e
$$
gives us
$$
2\alpha F-U\wedge U=0\ \Rightarrow\ F=\frac{1}{2\alpha}U\wedge U.
$$
That means
$$
S=\alpha e+U+\frac{1}{2\alpha}U\wedge U=\alpha\
\extexp(\frac{1}{\alpha}U).
$$
Denoting
$$
B=\frac{1}{\alpha}U,\quad \alpha=\pm\frac{1}{\sqrt{\lambda}},
$$
we get
$$
S=\pm\frac{1}{\sqrt{\lambda}}\extexp\,B,
$$
where
\begin{equation}
\lambda e=\extexp(B)\ \extexp(-B).\label{lam:B}
\end{equation}

It is easy to prove that if an element $S\in\Spin_+(p,q)$ is such
that $\Tr\,S=0$, then $S$ can be represented in the form
(\ref{S:Badj}). This completes the proof. \fin

If $S\in\Spin_+(p,q)$ has the form (\ref{exex}), then we say that
 $S$ is given with the aid of semi-polynomial parametrisation.
 That means, coefficients of $S$ in the basis (\ref{cl:basis})
are polynomials of second degree of coefficients $b_{ij}$
multiplied on one and the same factor $1/\sqrt{\lambda}$, where
$\lambda=\lambda(B)$ is the polynomial of second degree of
coefficients $b_{ij}$.

If $S\in\Spin_+(p,q)$ has the form (\ref{S:Badj}), then we say
that $S$ is given with the aid of adjoint semi-polynomial
parametrisation.

Let us write down explicit form of elements $S\in\Spin_+(p,q)$,
$p+q=4$ using real coefficients of $b_{ij}\in B\in\cl_2(p,q)$
$$
B= b_{12}\,e^{12} + b_{13}\,e^{13} + b_{14}\,e^{14} +
     b_{23}\,e^{23} + b_{24}\,e^{24} + b_{34}\,e^{34}.
$$
We have
\begin{eqnarray*}
\extexp(B) &=& e + b_{12}\,e^{12} + b_{13}\,e^{13} +
     b_{14}\,e^{14} + b_{23}\,e^{23} +\\
     && b_{24}\,e^{24} +
     b_{34}\,e^{34} + (b_{14}\,b_{23} - b_{13}\,b_{24} +
        b_{12}\,b_{34})\,e^{1234}
\end{eqnarray*}

Expressions for the scalar $\lambda=\lambda(B)$ that satisfy
(\ref{lam:B}) and for $\epsilon(1+\beta)$ are depend on a
signature $(p,q)$.

For $(p,q)=(0,4),(4,0)$
\begin{eqnarray*}
\lambda &=&  1 + b_{12}{}^2 + b_{13}{}^2 + b_{14}{}^2 +
b_{23}{}^2+
     b_{14}{}^2\,b_{23}{}^2 - 2\,b_{13}\,b_{14}\,b_{23}\,b_{24}
     +
     b_{24}{}^2 +\\ && b_{13}{}^2\,b_{24}{}^2 +
     2\,b_{12}\,b_{14}\,b_{23}\,b_{34} - 2\,b_{12}\,b_{13}\,b_{24}\,b_{34} +
     b_{34}{}^2 +
     b_{12}{}^2\,b_{34}{}^2,\\
\epsilon(1+\beta) &=& 1 - b_{12}{}^2 - b_{13}{}^2 - b_{14}{}^2 -
b_{23}{}^2 -
     b_{24}{}^2 - b_{34}{}^2.
\end{eqnarray*}

For $(p,q)=(1,3)$
\begin{eqnarray*}
\lambda &=& 1 - b_{12}{}^2 - b_{13}{}^2 - b_{14}{}^2 + b_{23}{}^2
-
     b_{14}{}^2\,b_{23}{}^2 + 2\,b_{13}\,b_{14}\,b_{23}\,b_{24}
     +
     b_{24}{}^2 -\\&& b_{13}{}^2\,b_{24}{}^2 -
     2\,b_{12}\,b_{14}\,b_{23}\,b_{34} + 2\,b_{12}\,b_{13}\,b_{24}\,b_{34} +
     b_{34}{}^2 -
     b_{12}{}^2\,b_{34}{}^2,\\
\epsilon(1+\beta) &=& -1 - b_{12}{}^2 - b_{13}{}^2 - b_{14}{}^2 +
b_{23}{}^2 +
     b_{24}{}^2 + b_{34}{}^2.
\end{eqnarray*}

For $(p,q)=(2,2)$
\begin{eqnarray*}
\lambda &=& 1 + b_{12}{}^2 - b_{13}{}^2 - b_{14}{}^2 - b_{23}{}^2
+
     b_{14}{}^2\,b_{23}{}^2 - 2\,b_{13}\,b_{14}\,b_{23}\,b_{24} -
     b_{24}{}^2 +\\&& b_{13}{}^2\,b_{24}{}^2 + 2\,b_{12}\,b_{14}\,b_{23}\,b_{34} -
     2\,b_{12}\,b_{13}\,b_{24}\,b_{34} + b_{34}{}^2 +
     b_{12}{}^2\,b_{34}{}^2,\\
\epsilon(1+\beta) &=& 1 - b_{12}{}^2 + b_{13}{}^2 + b_{14}{}^2 +
b_{23}{}^2 +
     b_{24}{}^2 - b_{34}{}^2.
\end{eqnarray*}

For $(p,q)=(3,1)$
\begin{eqnarray*}
\lambda &=& 1 + b_{12}{}^2 + b_{13}{}^2 - b_{14}{}^2 + b_{23}{}^2
-
     b_{14}{}^2\,b_{23}{}^2 + 2\,b_{13}\,b_{14}\,b_{23}\,b_{24} -
     b_{24}{}^2 -\\&& b_{13}{}^2\,b_{24}{}^2 - 2\,b_{12}\,b_{14}\,b_{23}\,b_{34} +
     2\,b_{12}\,b_{13}\,b_{24}\,b_{34} - b_{34}{}^2 -
     b_{12}{}^2\,b_{34}{}^2,\\
\epsilon(1+\beta) &=& -1 + b_{12}{}^2 + b_{13}{}^2 - b_{14}{}^2 +
b_{23}{}^2 -
     b_{24}{}^2 - b_{34}{}^2.
\end{eqnarray*}

With the aid of these formulas we get the general
form(\ref{exex}),(\ref{S:Badj}) of elements $S\in\Spin_+(p,q)$.

\medskip

The proof of the following  known theorem is straightforward.

\begin{theorem}
Let $n=p+q=2,3$. Any element $S\in\Spin_+(p,q)$ can be represented
in the following form
\begin{equation}
S=\pm e\sqrt{1+\beta}+B,\label{S:adj}
\end{equation}
where $B\in\cl_2(p,q)$ is such that $\beta=\Tr(B^2)\geq -1$. The
sign (plus or minus) at the right hand part of (\ref{S:adj}) is
equal to the sign of the number $\Tr\,S$. \fin
\end{theorem}

\medskip

\noindent{\bf Orthogonal groups.} Consider the Lie groups of
special orthogonal matrices of dimension $n=p+q$
$$
\SO(p,q)=\{P\in GL(n,\R) : P^T\eta P=\eta,\ det\,P=1\},
$$
where $P^T$ is the transposed matrix. For $pq\neq 0$ the groups
$\SO(p,q)$ have two disconnected components \cite{Diedonne}. The
component of the group $\SO(p,q)$ that contains the identity
matrix is a subgroup $\SO_+(p,q)$. In cases $pq=0$ we have
$\SO(0,n)=SO(n,0)=SO(n)$ and this group has only one component,
i.e. $\SO_+(n)=SO(n)$.

For $p+q=4$ the following propositions are valid \cite{Lounesto}:
\begin{itemize}
\item $\SO_+(4,0)\simeq\SO_+(0,4)\simeq\SO(4)$. \item
$\SO_+(1,3)\simeq\SO_+(3,1)$. \item $\forall P\in\SO(4)$ there
exists an anti-Hermitian matrix of fourth order ($A^T=-A$) such
that $P=\exp\,A$. \item $\forall P\in\SO_+(1,3)$ there exists a
matrix $A$ of fourth order such that $\eta A^T\eta=-A$,
($\eta=diag(1,-1,-1,-1)$) and $P=\exp\,A$. \item There exist
matrices from $\SO_+(2,2)$ that can't be represented in the form
$\pm\exp\,A$, where $\eta A^T\eta=-A$, ($\eta=diag(1,1,-1,-1)$).

\end{itemize}

\medskip

\noindent{\bf Connection between spinor and orthogonal groups.} It
is known \cite{Lounesto} that the spinor group $\Spin_+(p,q)$
double cover the orthogonal group $\SO_+(p,q)$. This connection
can be expressed by the formula
\begin{equation}
S^\sim e^a S=p^a_b e^b. \label{spin:ort}
\end{equation}
In this formula a pair of elements $\pm S$ of spinor group are
connected with the matrix $P=\|p^a_b\|$ from the orthogonal group.

\smallskip

Consider in more details the case  $(p,q)=(1,3)$, which is
important for physics. Let us take expressions for
$S\in\Spin_+(1,3)$ from Theorem 1 and, using (\ref{spin:ort}),
calculate corresponding elements of the matrix $P=\|p^a_b\|$. Then
we get the following formulas.

If $S$ has the form (\ref{exex}) and $\lambda=\lambda(B)>0$, then
elements of the matrix $T=\lambda\,P$ have the form
\begin{eqnarray*}
t^1_1 &=& 1 + b_{12}{}^2 + b_{13}{}^2 + b_{14}{}^2 + b_{23}{}^2 +
     b_{14}{}^2\,b_{23}{}^2 - 2\,b_{13}\,b_{14}\,b_{23}\,b_{24} +\\&&
     b_{24}{}^2 + b_{13}{}^2\,b_{24}{}^2 +
        2\,b_{12}\,b_{14}\,b_{23}\,b_{34} -
        2\,b_{12}\,b_{13}\,b_{24}\,b_{34} + b_{34}{}^2 +
     b_{12}{}^2\,b_{34}{}^2,\\
t^1_2 &=& 2\,b_{12} + 2\,b_{13}\,b_{23} + 2\,b_{14}\,b_{24} +
     2\,b_{14}\,b_{23}\,b_{34} - 2\,b_{13}\,b_{24}\,b_{34} +
     2\,b_{12}\,b_{34}{}^2,\\
t^1_3 &=& 2\,b_{13} - 2\,b_{12}\,b_{23} - 2\,b_{14}\,b_{23}\,
      b_{24} + 2\,b_{13}\,b_{24}{}^2 + 2\,b_{14}\,b_{34} -
     2\,b_{12}\,b_{24}\,b_{34},\\
t^1_4 &=& 2\,b_{14} + 2\,b_{14}\,b_{23}{}^2 - 2\,b_{12}\,b_{24} -
     2\,b_{13}\,b_{23}\,b_{24} - 2\,b_{13}\,b_{34} +
     2\,b_{12}\,b_{23}\,b_{34},\\
t^2_1 &=& 2\,b_{12} - 2\,b_{13}\,b_{23} - 2\,b_{14}\,b_{24} +
     2\,b_{14}\,b_{23}\,b_{34} - 2\,b_{13}\,b_{24}\,b_{34} +
     2\,b_{12}\,b_{34}{}^2,\\
t^2_2 &=& 1 + b_{12}{}^2 - b_{13}{}^2 - b_{14}{}^2 - b_{23}{}^2 +
     b_{14}{}^2\,b_{23}{}^2 - 2\,b_{13}\,b_{14}\,b_{23}\,b_{24} -\\&&
     b_{24}{}^2 + b_{13}{}^2\,b_{24}{}^2 +
        2\,b_{12}\,b_{14}\,b_{23}\,b_{34} -
        2\,b_{12}\,b_{13}\,b_{24}\,b_{34} + b_{34}{}^2 +
     b_{12}{}^2\,b_{34}{}^2,\\
t^2_3 &=& 2\,b_{12}\,b_{13} - 2\,b_{23} + 2\,b_{14}{}^2\,b_{23} -
     2\,b_{13}\,b_{14}\,b_{24} + 2\,b_{12}\,b_{14}\,b_{34} -
     2\,b_{24}\,b_{34},\\
t^2_4 &=& 2\,b_{12}\,b_{14} - 2\,b_{13}\,b_{14}\,b_{23} -
     2\,b_{24} + 2\,b_{13}{}^2\,b_{24} - 2\,b_{12}\,b_{13}\,b_{34} +
        2\,b_{23}\,b_{34},\\
t^3_1 &=& 2\,b_{13} + 2\,b_{12}\,b_{23} - 2\,b_{14}\,b_{23}\,b_{24}
+
        2\,b_{13}\,b_{24}{}^2 - 2\,b_{14}\,b_{34} -
     2\,b_{12}\,b_{24}\,b_{34},\\
t^3_2 &=& 2\,b_{12}\,b_{13} + 2\,b_{23} - 2\,b_{14}{}^2\,b_{23} +
     2\,b_{13}\,b_{14}\,b_{24} - 2\,b_{12}\,b_{14}\,b_{34} -
     2\,b_{24}\,b_{34},\\
t^3_3 &=& 1 - b_{12}{}^2 + b_{13}{}^2 - b_{14}{}^2 - b_{23}{}^2 +
     b_{14}{}^2\,b_{23}{}^2 - 2\,b_{13}\,b_{14}\,b_{23}\,b_{24} +\\&&
     b_{24}{}^2 + b_{13}{}^2\,b_{24}{}^2 +
        2\,b_{12}\,b_{14}\,b_{23}\,b_{34} -
        2\,b_{12}\,b_{13}\,b_{24}\,b_{34} - b_{34}{}^2 +
     b_{12}{}^2\,b_{34}{}^2,\\
t^3_4 &=& 2\,b_{13}\,b_{14} + 2\,b_{12}\,b_{14}\,b_{23} -
     2\,b_{12}\,b_{13}\,b_{24} - 2\,b_{23}\,b_{24} - 2\,b_{34} +
     2\,b_{12}{}^2\,b_{34},\\
t^4_1 &=& 2\,b_{14} + 2\,b_{14}\,b_{23}{}^2 + 2\,b_{12}\,b_{24} -
     2\,b_{13}\,b_{23}\,b_{24} + 2\,b_{13}\,b_{34} +
     2\,b_{12}\,b_{23}\,b_{34},\\
t^4_2 &=& 2\,b_{12}\,b_{14} + 2\,b_{13}\,b_{14}\,b_{23} +
     2\,b_{24} - 2\,b_{13}{}^2\,b_{24} + 2\,b_{12}\,b_{13}\,b_{34} +
        2\,b_{23}\,b_{34},\\
t^4_3 &=& 2\,b_{13}\,b_{14} - 2\,b_{12}\,b_{14}\,b_{23} +
     2\,b_{12}\,b_{13}\,b_{24} - 2\,b_{23}\,b_{24} + 2\,b_{34} -
     2\,b_{12}{}^2\,b_{34},\\
t^4_4 &=& 1 - b_{12}{}^2 - b_{13}{}^2 + b_{14}{}^2 + b_{23}{}^2 +
     b_{14}{}^2\,b_{23}{}^2 - 2\,b_{13}\,b_{14}\,b_{23}\,b_{24} -\\&&
     b_{24}{}^2 + b_{13}{}^2\,b_{24}{}^2 +
        2\,b_{12}\,b_{14}\,b_{23}\,b_{34} -
        2\,b_{12}\,b_{13}\,b_{24}\,b_{34} - b_{34}{}^2 +
     b_{12}{}^2\,b_{34}{}^2.
\end{eqnarray*}

If $S$ has the form (\ref{S:Badj}), $\epsilon(1+\beta)\geq 0$ and
$B\wedge B=0$, then elements of the matrix $P$ have the form
\begin{eqnarray*}
p^1_1 &=& -1 + 2\,b_{23}{}^2 + 2\,b_{24}{}^2 + 2\,b_{34}{}^2,\\
p^1_2 &=& 2\,b_{13}\,b_{23} + 2\,b_{14}\,b_{24} +
     2\,b_{34}\,\sqrt{\rho},\\
p^1_3 &=& -2\,b_{12}\,b_{23} + 2\,b_{14}\,b_{34} -
     2\,b_{24}\,\sqrt{\rho},\\
p^1_4 &=& -2\,b_{12}\,b_{24} - 2\,b_{13}\,b_{34} +
     2\,b_{23}\,\sqrt{\rho},\\
p^2_1 &=& -2\,b_{13}\,b_{23} - 2\,b_{14}\,b_{24} +
     2\,b_{34}\,\sqrt{\rho},\\
p^2_2 &=& -1 - 2\,b_{13}{}^2 - 2\,b_{14}{}^2 + 2\,b_{34}{}^2,\\
p^2_3 &=& 2\,b_{12}\,b_{13} - 2\,b_{24}\,b_{34} +
     2\,b_{14}\,\sqrt{\rho},\\
p^2_4 &=& 2\,b_{12}\,b_{14} + 2\,b_{23}\,b_{34} -
     2\,b_{13}\,\sqrt{\rho},\\
p^3_1 &=& 2\,b_{12}\,b_{23} - 2\,b_{14}\,b_{34} -
     2\,b_{24}\,\sqrt{\rho},\\
p^3_2 &=& 2\,b_{12}\,b_{13} - 2\,b_{24}\,b_{34} -
     2\,b_{14}\,\sqrt{\rho},\\
p^3_3 &=& -1 - 2\,b_{12}{}^2 - 2\,b_{14}{}^2 + 2\,b_{24}{}^2,\\
p^3_4 &=& 2\,b_{13}\,b_{14} - 2\,b_{23}\,b_{24} +
     2\,b_{12}\,\sqrt{\rho},\\
p^4_1 &=& 2\,b_{12}\,b_{24} + 2\,b_{13}\,b_{34} +
     2\,b_{23}\,\sqrt{\rho},\\
p^4_2 &=& 2\,b_{12}\,b_{14} + 2\,b_{23}\,b_{34} +
     2\,b_{13}\,\sqrt{\rho},\\
p^4_3 &=& 2\,b_{13}\,b_{14} - 2\,b_{23}\,b_{24} -
     2\,b_{12}\,\sqrt{\rho},\\
p^4_4 &=& -1 - 2\,b_{12}{}^2 - 2\,b_{13}{}^2 + 2\,b_{23}{}^2,
\end{eqnarray*}
where $$ \rho =\epsilon(1+\beta)= -1 - b_{12}{}^2 - b_{13}{}^2 -
b_{14}{}^2 +
        b_{23}{}^2 + b_{24}{}^2 + b_{34}{}^2\geq 0.
$$

\end{document}